\documentstyle[preprint,aps]{revtex}
\begin{document}

\title{Electron transport in  Coulomb- and tunnel-coupled one-dimensional
systems}

\author{ O. E. Raichev\cite{oleg} and P. Vasilopoulos\cite{takis}%\\
\ \\}
\voffset 1cm
\address{ \cite{oleg}Institute of Semiconductor Physics, NAS Ukraine,
Pr. Nauki 45, Kiev, 252650, Ukraine
\ \\
%\ \\
\cite{takis}Concordia University, Department of Physics, Montr\'{e}al,
Qu\'{e}bec, Canada, H3G 1M8}

\date{\today}

\maketitle

\begin{abstract}

We develop a linear theory of electron transport for a system of two
identical quantum wires in a wide range of the wire length $L$,
unifying both the ballistic and diffusive transport regimes. The
microscopic model, involving the interaction of electrons with each
other and with bulk acoustical phonons allows a reduction of the
quantum kinetic equation to a set of coupled equations for the
local chemical potentials for forward- and backward-moving electrons
in the wires. As an application of the general solution
of these equations, we consider different kinds of electrical
contacts to the double-wire system and calculate the direct
resistance, the transresistance, in the presence of tunneling and
Coulomb drag, and the tunneling resistance.
If $L$ is smaller than the backscattering length $l_P$, both the
tunneling and the drag lead to a negative transresistance, while in
the diffusive regime ($L \gg l_P$) the tunneling opposes the drag
and leads to a positive transresistance. If $L$ is smaller than
the phase-breaking length, the tunneling leads to interference
oscillations of the resistances that are damped exponentially
with $L$.\\


\end{abstract}

\pagebreak

\section{Introduction}

One-dimensional (1D) electron systems, such as they occur in semiconductor
quantum wires, are in the forefront of research in modern
condensed-matter physics. In submicrometer-long quantum wires at low
temperatures the electron transport occurs in the ballistic regime$^1$
and the wire conductance reaches its fundamental value of
$G_0=e^2/\pi \hbar$. On the other hand, in sufficiently long wires the
conductance is limited by scattering processes. If quantum-interference
effects are neglected, as is the case when the inelastic scattering
dominates, the conductance is given by $\sigma/L$, where $L$ is the wire
length and $\sigma$ the conductivity described by the Drude expression
$\sigma=e^2 n \tau_{tr}/m$, where $n$ is the electron density, $\tau_{tr}$
the transport time, and $m$ the effective mass of the electron. This
regime is referred to as the diffusive transport.

Modern technology allows to create various systems comprising two
quantum wires put closely to each other so that the tunneling of
electrons between the wires and/or interlayer electron-electron
interaction is essential. Both these effects give rise to coupling
between the electron sub-systems in single wires and in that way
modify their electronic properties. This renders the coupled
double-wire systems a subject of interest. In the  past years, there
have been experimental and theoretical studies of 1D-1D tunneling$^{2,3}$
and electron transport$^{4-22}$ along the wires of such
systems. Investigations of the transport are mostly devoted to
interlayer tunneling in the purely ballistic regime and in connection with
the idea of electron-wave coupler$^{4,5}$. On the other hand, there
are theoretical papers$^{18-22}$ describing the momentum transfer
between the wires due to interlayer Coulomb interaction and the
corresponding interlayer transresistance (Coulomb drag).
Calculations of the Coulomb drag have been done both for the
diffusive$^{18-20}$ and ballistic$^{21}$ transport regimes, as well as for
the regime in which the electron sub-systems are described by the Luttinger
liquid model$^{22}$.

Despite this progress, there is a substantial lack of description of the
electron transport in coupled quantum wires. Even if we accept
the concept that the electrons are described by a normal Fermi liquid,
two important questions arise. The first one is how to describe the
electrical properties when both tunneling and interactions of
the electrons (with each other and with impurities or phonons)
are essential. The second one is how to bridge the gap between
the ballistic and diffusive transport regimes in such a description.

In this paper we present a linear-response theory of electron transport
in coupled quantum wires that gives a reasonable answer to both
questions stated above. We consider two parallel, tunnel-coupled
1D systems of degenerate electron gases adiabatically contacted to four
equilibrium reservoirs, as shown and labeled in Fig. 1. This general
scheme of a four-terminal device may describe both planar$^{2,4,5,15}$
and vertically coupled$^{3,9,12,17}$ double-wire devices. We take into
account the interaction of electrons with themselves as well as that with
acoustical phonons. We start from the quantum kinetic equation and finally
transform it to a set of linear differential equations describing the
distributions of the local chemical potentials for the systems of forward-
and backward-moving electrons along the wires. The boundary conditions for
such equations are dictated by the Landauer-B{\"u}ttiker-Imry theory.
This transformation is justified from microscopic calculations, which
also give us expressions for the characteristic times,
associated with the interactions involved, that enter the
equations for chemical potentials.

As an application of our transport theeory, we analyze in detail
different kinds of electrical contacts to the double-wire system.
First we consider the case when the voltage is applied between the ends of
one wire and calculate the "direct" resistance of this wire as it is
affected by the presence of the other one, as well as the transresistance,
i.e., the resistance associated with the voltage induced at the ends of the
uncontacted wire. 
%A
Details about  experimental 
measurements of the transresistance in such systems can be found 
in Ref. 23. 
%A
Next we consider the case corresponding to
the tuneling measurements$^2$, when the voltage is applied between the
wires, and calculate the tunneling resistance. Several previously
obtained theoretical results for such quantities (some of them are for
coupled 2D systems) follow from our theory as limiting cases. A brief
account of the main results appeared in Ref. 23.

The paper is organized as follows. In Sec. II we present the microscopic
model and derive the equations for the local chemical potentials in the
layers. In Sec. III we solve these equations, apply the obtained
results to the calculation of the direct resistance, transresistance,
and tunneling resistance of the double-wire system, and describe possible
transport regimes. Concluding remarks and discussion of the approximations
made are given in Sec. IV. The Appendix contains detailed microscopic
calculations and expressions of the characteristic times entering the
equations for chemical potentials.

\section{From quantum kinetic theory to local description}

Consider two homogeneous 1D quantum layers of length $L$, labeled
left ($l$) and right ($r$) along the $x$ axis, see Fig. 1.
The quantum kinetic equation for the density matrix $\hat{\rho}$ reads

\begin{equation}
\partial \hat{\rho}/ \partial t + (i/\hbar) [\hat{H}_0 +
\hat{H}_C + \hat{H}_{e-ph}, \hat{\rho}] =0.
%1
\end{equation}
Here we assume that electrons interact with each other via the Coulomb
field $\hat{H}_C$ and with acoustical phonons $\hat{H}_{e-ph}$.
Elastic scattering is neglected, i.e., we assume ideal wires. The
unperturbed Hamiltonian $\hat{H}_0$ includes both the kinetic and
potential energy operators. Below we use the basis of the isolated
$l$ and $r$ layer states $F_l(y,z)$ and $F_r(y,z)$ and assume that
only the lowest level is occupied in each layer. In this basis the
potential energy is the matrix

\begin{equation}
\hat{h}=(\Delta/2) \hat{\sigma}_z + T \hat{\sigma}_x.
%2
\end{equation}
Here $\hat{\sigma}_i$ are the Pauli matrices, $\Delta$ is the
level-splitting energy, and $T$ the tunneling matrix element
characterizing the strength of the tunnel coupling. Such tight-binding
description is often used in application to two-level systems.%$^{23}$.

The kinetic equation can be written$^{25}$ as one
for the Keldysh's Green's function $\hat{G}^{-+}$. Below we
consider the case when the characteristic spatial scale of
the electronic distribution is large in comparison to the Fermi wavelength
$\pi \hbar/p_F$ and use the Keldysh's matrix Green's function in the
Wigner representation $\hat{G}^{-+}_{{\textstyle \varepsilon, t}}(p,x)$,
where $p$ and $\varepsilon$ are the momentum and energy and $t$
the time. The time dependence of $\hat{G}^{-+}$ is not essential in
the following, since we study a time-averaged, steady-state
problem. The linear response theory uses a Green's function of the form

        \begin{equation}
        \hat{G}^{\alpha \beta}_{{\textstyle \varepsilon}}(p,x)=
        \hat{G}^{(0) \alpha \beta}_{{\textstyle \varepsilon}}(p) + \delta
        \hat{G}^{\alpha \beta}_{{\textstyle \varepsilon}}(p,x),
        %3
        \end{equation}
where $\alpha$ and $\beta$ are $+$ or $-$. The unperturbed part
$\hat{G}^{(0) \alpha \beta}_{{\textstyle \varepsilon}}(p)$ is given by

        \begin{equation}
        \hat{G}^{(0)-+}_{{\textstyle \varepsilon}}(p)= f(\varepsilon)
        [\hat{G}^{A}_{{\textstyle \varepsilon}}(p)
        -\hat{G}^{R}_{{\textstyle
        \varepsilon}}(p)],~~\hat{G}^{(0)+-}_{{\textstyle \varepsilon}}(p)=
        \left( f(\varepsilon)-1 \right)
        [\hat{G}^{A}_{{\textstyle \varepsilon}}(p)
        -\hat{G}^{R}_{{\textstyle
        \varepsilon}}(p)],
        %4
        \end{equation}
        \begin{equation}
        \hat{G}^{(0)--}_{{\textstyle \varepsilon}}(p)=
        \hat{G}^{R}_{{\textstyle \varepsilon}}(p)+
        \hat{G}^{(0)-+}_{{\textstyle
        \varepsilon}}(p), ~~\hat{G}^{(0)++}_{{\textstyle \varepsilon}}(p)=
        -\hat{G}^{A}_{{\textstyle \varepsilon}}(p)+
        \hat{G}^{(0)-+}_{{\textstyle
        \varepsilon}}(p),
        %5
        \end{equation}
where $f(\varepsilon)=[1+e^{(\varepsilon-\mu)/k_B T_e}]$ is the
equilibrium Fermi distribution function and $\hat{G}^{R,A}$ are the
retarded and advanced Green's functions which satisfy the equations

        \begin{equation}
        \left[ \varepsilon- \varepsilon_p - \hat{h} -
        \hat{\Sigma}^{R,A}_{{\textstyle \varepsilon}}(p) \right]
        \hat{G}^{R,A}_{{\textstyle \varepsilon}}(p)=1;
        %6
        \end{equation}
here $\hat{\Sigma}^{R,A}_{{\textstyle \varepsilon}}(p)$ are the
self-energy functions.

The linearized kinetic equation reads

        \begin{eqnarray}
        \frac{\hbar}{2} \left\{ \hat{v}_p, \frac{\partial}{\partial x}
        \delta \hat{G}^{-+}_{{\textstyle \varepsilon}}(p,x) \right\}
        + i \left[\hat{h}, \delta \hat{G}^{-+}_{{\textstyle
        \varepsilon}}(p,x) \right] \nonumber \\
        - \frac{\hbar}{2} \left\{ \frac{\partial}{\partial x}
        \hat{\varphi} , \frac{\partial}{\partial p}
        \hat{G}^{(0)-+}_{{\textstyle \varepsilon}}(p) \right\}
        + i \left[\hat{\varphi}, \hat{G}^{(0)-+}_{{\textstyle
        \varepsilon}}(p) \right]
        = i \delta \hat{{\cal I}}(\varepsilon,p,x).
%7
\end{eqnarray}
Here $\{...\}$ denotes anticommutators, $\hat{v}_p=\hat{P}_l
v_{lp}+ \hat{P}_r v_{rp}$ is the diagonal matrix of the group velocities,
and $\hat{P}_l=(1+\hat{\sigma}_z)/2$ and $\hat{P}_r=(1-\hat{\sigma}_z)/2$
are the projection matrices. In this paper we consider the case of equal
group velocities in the layers, with $v_{lp}=v_{rp}=v_p=p/m$.
If a magnetic field is applied perpendicular to the wire plane
though, $v_{lp}$
and $v_{rp}$ become different. Further, $\hat{\varphi}$ is the matrix of
the self-consistent electrostatic potential arising due to perturbation
of the electron density. In the mean-field (Hartree) approximation this
matrix is diagonal. Finally, the generalized collision integral
$\hat{{\cal I}}$ is given by$^{25}$

        \begin{equation}
        \hat{{\cal I}}= - \left\{ \hat{\Sigma}^{-+} \hat{G}^{++} +
        \hat{\Sigma}^{--} \hat{G}^{-+} + \hat{G}^{-+} \hat{\Sigma}^{++} +
        \hat{G}^{--} \hat{\Sigma}^{-+} \right\},
        %8
        \end{equation}
where all Green's functions $\hat{G}^{\alpha \beta}$ and self-energy
functions $\hat{\Sigma}^{\alpha \beta}$ have the same arguments $\varepsilon$,
$p$, and $x$. This corresponds to a quasiclassical description of
the scattering. However, the matrix structure of $\hat{G}^{\alpha \beta}$
and $\hat{\Sigma}^{\alpha \beta}$ remains important and Eq. (7) is
not reduced to a classical Boltzmann equation. Since we consider
the interaction of electrons with each other and with acoustical
phonons, the corresponding lowest-order contributions to the self-energy
are given by the diagrams of Fig. 2. We neglect the exchange part
of the Coulomb interaction for the following reasons. The first-order
exchange contributions do not influence the imaginary part of the
self-energy and are not, therefore, essential for the calculation of the
collision integral. The second-order exchange contributions are small
as compared to the second-order direct Coulomb contributions,
represented by the diagram of Fig. 2 (b), if the momentum transfer $q$ is
small in comparison with the Fermi momentum. Finally, there is no exchange
contributions to the interlayer Coulomb interaction.

We consider low temperatures and degenerate electrons. We
also assume that the Fermi energy is large in comparison with both the
tunneling matrix element $T$ and the level splitting $\Delta$, thereby
neglecting the difference between the electron densities in the layers.
We sum up Eq. (7) over the electron momentum $p$ in the regions
of positive ($+$) (or forward) and  negative ($-$) (or backward )
group velocities and introduce the nonequilibrium part
$\hat{g}_{\textstyle \varepsilon}(x)$ of the energy
distribution function in  the manner

        \begin{equation}
        \hat{g}^{\pm}_{\textstyle \varepsilon}(x)= \int_{\pm}
        \frac{d p}{2\pi i} |v_p| \delta \hat{G}^{-+}_{{\textstyle
        \varepsilon}}(p,x).
        %9
        \end{equation}
Since $\delta \hat{G}^{-+}$ is essentially nonzero only in narrow intervals
of energy and momentum near the equilibrium chemical potential $\mu$ and
Fermi momentum $p_F$, we can replace $|v_p|$ in this equation by the Fermi
velocity $v_F$, common to both layers. The integration in the $+$ and $-$
regions in Eq. (7) removes the contributions proportional to the potential
matrix $\hat{\varphi}(x)$ and we obtain

        \begin{equation}
        \pm v_F\frac{\partial}{\partial x} \hat{g}^{\pm}_{\textstyle
        \varepsilon}(x)
        + \frac{i}{\hbar}
        \left[\hat{h}, \hat{g}^{\pm}_{\textstyle \varepsilon}(x) \right]
        =  \delta \hat{I}_{\pm}(\varepsilon,x)
        %10
        \end{equation}
where the collision integral $\delta \hat{I}_{\pm}(\varepsilon,x)=
(2 \pi \hbar)^{-1} \int_{\pm} d p |v_p|~\delta
\hat{{\cal I}}(\varepsilon, p,x)$ depends on both $\hat{g}^{+}$
and $\hat{g}^{-}$, since it accounts for both %c
forward and backward scattering processes. However,
when we integrate Eq. (10) over the
energy, the diagonal part of $\delta \hat{I}_{\pm}(\varepsilon,x)$
vanishes for forward-scattering contributions, and only the backscattering
contributions remain, see below. In contrast, the forward-scattering
contributions for the nondiagonal part of the collision integral are not 
eliminated by the energy integration.

The matrix kinetic equation (10) is equivalent to eight scalar equations
for the four components of $\hat{g}^{+}$ and the four ones %components
of $\hat{g}^{-}$, corresponding to forward- and backward-propagating
electrons, respectively. These equations must be accompanied by boundary
conditions connecting the components of $\hat{g}^{\pm}$ with the
quasi-equilibrium distribution functions of the four leads which
the quantum wires are contacted to, cf.  Fig. 1. The distribution
functions of the leads are defined by the four chemical potentials
$\mu_{1l}$, $\mu_{1r}$, $\mu_{2l}$, and $\mu_{2r}$. If we assume
that the potentials in the contact regions are sufficiently smooth in
comparison with the Fermi wavelength but abrupt enough as compared to
the characteristic scale of the electronic distribution, we can apply
Eq. (10) in the contact region as well. It gives us the conditions of
continuity for all components of $\hat{g}^{\pm}_{\textstyle
\varepsilon}$(x) across the contact regions and we obtain

        \begin{eqnarray}
        \hat{g}^{+}_{\textstyle \varepsilon}(0)=
        - \frac{ \partial f(\varepsilon)}{\partial
        \varepsilon} [\hat{P}_l \delta \mu_{1l} +
        \hat{P}_r \delta \mu_{1r}],~~
        \hat{g}^{-}_{\textstyle \varepsilon}(L)=
        - \frac{ \partial f(\varepsilon)}{\partial
        \varepsilon} [\hat{P}_l \delta \mu_{2l} +
        \hat{P}_r \delta \mu_{2r} ],
        %11
        \end{eqnarray}
with $\delta \mu_{1l}=\mu_{1l} - \mu$, etc.
The forward- and backward-propagating states are "connected",
respectively, to the leads 1 and 2. The nondiagonal components
vanish at the contacts because the tunneling is absent
outside the region $x=[0,L]$.

The problem described by the matrix equation (10) and the
boundary conditions (11) can be considerably simplified and
solved analytically if we assume that both backscattering and the
interlayer tunneling occur much less frequently than the scattering of
electrons inside the layers and inside the $+$ or $-$ regions. The
tunneling can be made weak if, for example, the potential
barrier between the wires is thick enough. As concerns
the backscattering, this condition is often fulfilled at low temperatures
for both the electron-electron and electron-phonon scattering mechanisms.
In the first case, the backscattering probability contains a factor
$[K_0(2 p_F a)]^2$, where $K_0$ is the modified Bessel function and
$a$ is the wire width. This factor is exponentially small for
$2 p_F a >1$. The acoustic phonon-assisted backscattering gives a
small contribution in comparison with the electron-electron
forward-scattering due to the smallness of the electron-phonon
coupling constant. In addition, this backscattering is suppressed
at very low temperatures $T_e < 2 p_F s$, where $s$ is the sound
velocity. If the stated conditions are fulfilled, the diagonal
part of the energy distribution function of the electrons
have a Fermi-like energy dependence because any quasi-equilibrium
Fermi function locally satisfies a kinetic equation with the
electron-electron and electron-phonon collision integrals. It means that 
the diagonal part of $\hat{g}^{\pm}_{\textstyle \varepsilon}(x)$
is given by the following equation

        \begin{equation}
        [\hat{g}^{\pm}_{\textstyle \varepsilon}(x) ]_{jj}=
        - \frac{ \partial f(\varepsilon)}{\partial \varepsilon}
        \ \delta \mu_j^{\pm}(x)
        %12
        \end{equation}
where $\delta \mu_j^{\pm}(x)=\mu_j^{\pm}(x)-\mu$ ($j=l,r$) do not depend
on the energy. The quantities $\mu_j^{\pm}(x)$ have the direct meaning of
{\it local chemical potentials} for the layers $l$ and $r$. It is
convenient to introduce also the nondiagonal components of the chemical
potentials $\mu_{lr}^{\pm}(x) = \mu_u^{\pm}(x) - i \mu_v^{\pm}(x)$ and
$\mu_{rl}^{\pm}(x) = \mu_u^{\pm}(x) + i \mu_v^{\pm}(x)$, by writing the
whole chemical potential matrix as

        \begin{equation}
        \delta \hat{\mu}^{\pm}(x)=\int d
        \varepsilon ~\hat{g}^{\pm}_{\textstyle \varepsilon}(x).
        %13
        \end{equation}
Below we drop the symbol "$\delta$" in $\delta \hat{\mu}^{\pm}(x)$ and the
contact potentials $\delta \mu_{1l,r}$ and $\delta \mu_{2l,r}$ since
all chemical potentials are counted from the same equilibrium value $\mu$.

Substituting Eq. (12) into Eq. (10) and integrating the latter over the
energy, we finally obtain eight coupled, first-order differential
equations for the eight components of $\mu_k^{\pm}(x)$:

        \begin{eqnarray}
        \pm d \mu_l^{\pm}/dx + (\mu_l^{\pm} -\mu_l^{\mp})(1/l_P+1/l_D) -
        (\mu_r^{\pm}
        -\mu_r^{\mp})/l_D - 2t_F  \mu_v^{\pm} =0, \\
        \pm d \mu_r^{\pm}/dx + (\mu_r^{\pm} -\mu_r^{\mp})(1/l_P+1/l_D) -
        (\mu_l^{\pm}
        -\mu_l^{\mp})/l_D + 2t_F  \mu_v^{\pm} =0, \\
        \pm d \mu_u^{\pm}/dx + \delta_F \mu_v^{\pm} + \mu_u^{\pm}/l_C=0,
\\
        \pm d \mu_v^{\pm}/dx - \delta_F \mu_u^{\pm} + \mu_v^{\pm}/l_C +
        t_F  (\mu_l^{\pm} -\mu_r^{\pm}) =0.
        %C 13-16
        %14-17
        \end{eqnarray}
Here $t_F= T/\hbar v_F$ and $\delta_F=\Delta/\hbar v_F$. The boundary
conditions for all potentials follow from Eqs. (11)-(13) and are
$\mu_l^+(0)=\mu_{1l}$, $\mu_l^-(L)=\mu_{2l}$, $\mu_r^+(0)=\mu_{1r}$,
$\mu_r^-(L)=\mu_{2r}$ and $\mu_{u,v}^+(0)=\mu_{u,v}^-(L)=0$.
The characteristic lengths $l_P$, $l_D$, and $l_C$ result from
the collision integral $\delta \hat{I}_{\pm}(\varepsilon,x)$,
evaluated to the lowest order with respect to the tunneling matrix
element $T$, see Appendix for details. They are expressed, respectively,
through the phonon-assisted 1D transport time$^{26}$ $\tau_P$, the 1D
Coulomb drag time$^{19,20}$ $\tau_D$, and the phase-breaking time $\tau_C$
describing the suppression of tunnel coherence, as $l_P = 2 v_F \tau_P$,
$l_D = 2 v_F \tau_D$, and $l_C = v_F \tau_C$. The transport time
$\tau_P$ is common to both layers since we assume that the confining
potentials for the wires $l$ and $r$ are almost identical. The analytical
expressions for the $\tau_P$, $\tau_D$ and $\tau_C$ are given in the
Appendix. All  characteristic lengths are sensitive to the temperature $T$
and the level splitting $\Delta$. It is essential that $l_C$, which is
controlled by electron-electron interaction, is always much smaller than
$l_D$ and $l_P$. On the other hand, depending on the temperature and level 
splitting one can have different relations between $l_D$ and $l_P$: both cases,
$l_P \ll l_D$ and $l_P \gg l_D$, are possible.

Equations (14)-(17) with the stated boundary conditions give us a complete
description of the electrical properties of double quantum-wire systems
in a wide range of regimes starting from the purely ballistic transport
regime $L \ll l_C$ to the diffusive transport regime $L \gg l_P, l_D$.
The local currents flowing in the layers $j=l,r$ are expressed by

        \begin{equation}
        J_{j}(x)= G_0 [ \mu_j^+(x) - \mu_j^-(x) ]/e,
        %18
        \end{equation}
and the local tunnel currents are proportional to $T \mu_v^{\pm}(x)$.

Below we present the general solution of Eqs. (14)-(17) and
describe two important cases, that of {\it long} systems, with $L \gg
l_C$, and
that of {\it short} systems with $L \sim l_C$. To characterize the effects
of drag and tunneling, we then consider different kinds of electrical
contact to the double-wire system. First we consider a typical
setup for the drag measurements, when the current $J_r=J_r(0)=J_r(L)$ is
injected in wire $r$ ("drive wire") while no current is allowed to
flow into wire $l$, $J_l(0)=J_l(L)=0$, and calculate the transresistance
$R_{TR}$ defined as $R_{TR}=[\mu_{1l}- \mu_{2l}]/eJ_r$ as well as the
"direct" resistance $R=[\mu_{1r}-\mu_{2r}]/eJ_r$. Next, we turn to
the tunneling measurements$^2$, when the voltage is applied between the
wires. We consider both the symmetric setup, when all four ends of the
wires are connected to external sources, with $\mu_{1l}=\mu_{2l}$ and
$\mu_{1r}=\mu_{2r}$, and the non-symmetric one, when the voltage is
applied between the ends $1l$ and $2r$ while the remaining ends are
not contacted, $J_r(0)=0$, $J_l(L)=0$. For each of these cases we
calculate the tunneling resistances $R_{Ts}$ (symmetric) and $R_{Tn}$
(non-symmetric). Both of them can be defined as $[\mu_{1l}-\mu_{2r}]/
eJ_{T}$, where $J_{T}$ is the total current injected. $J_{T}$  is equal
to $2 J_l(0)$ and $J_l(0)$ for the symmetric and non-symmetric
contacts, respectively.

\section{Results}

Since Eqs. (14)-(17) are linear, their general solution  is easily obtained
as

        \begin{eqnarray}
        2\mu_{l,r}^+(x) =
        (1+L/l_P)^{-1} \left[ ( \mu_{1l}+ \mu_{1r} )
        \left( 1+ (L-x)/l_P \right) + ( \mu_{2l}+ \mu_{2r} )
        x/l_P \right] \nonumber \\
        \pm \sum_{i} \left( A_i^+ e^{ \lambda_i x} +  B_i^+ e^{-
        \lambda_i x}
        \right) \\
        2\mu_{l,r}^-(x) =
        (1+L/l_P)^{-1} \left[ (\mu_{1l}+ \mu_{1r} )
        (L-x)/l_P  + ( \mu_{2l}+ \mu_{2r} )\left( 1 +
        x/l_P \right) \right] \nonumber \\
        \pm \sum_{i} \left( A_i^- e^{ \lambda_i x} +  B_i^- e^{-
        \lambda_i x}
        \right) \\
        \mu_{v}^{\pm}(x) =
        \sum_{i} \left( C_i^{\pm} e^{ \lambda_i x} +  D_i^{\pm} e^{-
        \lambda_i x}
        \right) \\
        \mu_{u}^{\pm}(x) =
        - \sum_{i} \left( \frac{\delta_F}{\pm \lambda_i+ l_C^{-1}}
        C_i^{\pm} e^{ \lambda_i x} + \frac{\delta_F}{\mp \lambda_i +
        l_C^{-1}} D_i^{\pm} e^{- \lambda_i x} \right).
        %19-22
        \end{eqnarray}
Here $\lambda_i=\sqrt{y_i}$ and $y=y_i$ are the solutions of the cubic equation

        \begin{eqnarray}
        y^3-2 y^2 [l_C^{-2}- \delta_F^2 -4 t_F^2 ] +y [l_C^{-4}+
        (\delta_F^2
        +4 t_F^2)^2
        %\nonumber \\
        -8 (t_F/ l_C)^2 + 2 (\delta_F/l_C)^2 + 8 t_F^2/l_C l_1  ]
        \nonumber \\
        -4 t_F ^2 [4(t_F /l_C)^2 +(2/l_C l_1) (l_C^{-2}+ \delta_F^2) ]=0,
        %23
        \end{eqnarray}
where $l_1^{-1}=l_P^{-1}+2 l_D^{-1}$. The coefficients $A_i^{\pm}$,
$B_i^{\pm}$, $C_i^{\pm}$, and $D_i^{\pm}$ are to be found from Eqs.
(14)-(17) and the relevant boundary conditions. Below we use the property
$l_C \ll l_P,l_D$
and the condition of weak tunnel coupling $t_F \ll l_C^{-1}$ to simplify
this procedure. Then the three roots of Eq. (23) are easily obtained as

        \begin{equation}
        \lambda_1=\lambda \simeq 2(1/l_T^2+1/l_T l_1)^{1/2},  \ \
        \lambda_{2,3}= \lambda_{\pm} \simeq 1/l_C \pm i  \delta_F,
        %24
        \end{equation}
where we introduced the tunneling length $l_T=v_F \tau_T$.
The tunneling time $\tau_T$, which contains a resonance dependence
on the level splitting, is defined by

        \begin{equation}
        \tau_T^{-1}= \tau_C^{-1} \frac{2T^2}{\Delta^2+ (\hbar/\tau_C)^2 }.
        %25
        \end{equation}
The root $\lambda$ describes long-scale variations of the chemical
potentials while $\lambda_{\pm}$ corresponds to short-scale variations.
Accordingly, we consider the regimes that follow.

\subsection{Long wires, $L \gg l_C$.}

This length range comprises the region from the
"pseudo-ballistic" ($l_C \ll L \ll l_P,l_D$)
to the diffusive ($L \gg l_P$) regimes. All
solutions containing $\lambda_{\pm}$ exist only in short regions in
the vicinity of the contacts. They are evanescent inside the wire
region and not essential in the calculation of the currents.
Considering only the solutions involving $\lambda$, we find

        \begin{eqnarray}
        2 \mu_{l,r}^+(x) =
        (1+L/l_P)^{-1} \left[ (\mu_{1l}+ \mu_{1r} )
        \left( 1+ (L-x)/l_P \right) + (\mu_{2l}+ \mu_{2r} )
        x/l_P \right] \nonumber \\
        \pm ( \mu_{1l} - \mu_{1r} ) P(L-x)/P(L) %\left[\left(1+2l_1/l_T
        %\right) \sinh \lambda x_{L}+ \lambda l_1 \cosh \lambda x_{L}\right]
        \pm (\mu_{2l} - \mu_{2r} ) \sinh \lambda x /P(L)\\
        \nonumber \\
        2 \mu_{l,r}^-(x) =
        (1+L/l_P)^{-1}\left[ ( \mu_{1l}+ \mu_{1r} )
        (L-x)/l_P  +( \mu_{2l}+ \mu_{2r} ) \left( 1 +
        x/l_P \right) \right] \nonumber \\
        \pm (\mu_{1l} - \mu_{1r} ) \sinh \lambda(L-x)/P(L)
        \pm ( \mu_{2l} - \mu_{2r} )P(x)/P(L),
        %26,27
        \end{eqnarray}
where $P(x)= (1 + 2 l_1/l_T) \sinh \lambda x + \lambda l_1 \cosh \lambda x$.
The same expressions can be obtained from the four coupled balance
equations compactly presented as
        \begin{eqnarray}
        \pm d \mu_j^{\pm}/dx + (\mu_j^{\pm}-\mu_j^{\mp})(1/l_P+1/l_D)
        \nonumber \\
        - (\mu_{j'}^{\pm} -\mu_{j'}^{\mp})/l_D + (\mu_{j}^{\pm}-
        \mu_{j'}^{\pm})/l_T =0,
        %28
        \end{eqnarray}
where $j=l,r$ and $j' \neq j$. These equations follow from
Eqs. (14)-(17) in the limit $L \gg l_C$, when one can neglect
the derivatives $d \mu^{\pm}_{u,v}/ d x$ in comparison to
$\mu^{\pm}_{u,v}/l_C$. With Eqs. (26) and (27) we obtain

        \begin{equation}
        R = {\pi \hbar\over 2e^2}\left[ 2 +  L/l_P +
        \left(1+l_T/l_1 \right)^{1/2} \tanh(\lambda L/ 2) \right],
        %29
        \end{equation}
        \begin{equation}
        R_{TR}= {\pi \hbar\over 2e^2}\left[L/l_P -
        \left(1+l_T/ l_1 \right)^{1/2} \tanh(\lambda L/2)\right],
        %30
        \end{equation}

For $L \ll l_P, l_T$ we have $R \simeq \pi \hbar/e^2=G_0^{-1}$, while
the transresistance is given by $R_{TR} \simeq -(\pi \hbar/e^2)L [1/l_D +
1/2 l_T]$. As seen, $R_{TR}$ is small, always negative, and proportional
to the wire length $L$ multiplied by a sum of drag and tunneling rates.
If one neglects tunneling, the resulting expression for $R_{TR}$,
with $\tau_D$ given by Eq. (A11), describes the {\it Coulomb drag in the
ballistic regime} previously investigated$^{21}$ by Gurevich {\em et al.}.
When $L$ increases and the electron transport becomes diffusive
($L \gg l_P$), we obtain, for $\lambda L/2 \ll 1$,
$R \simeq (\pi \hbar /e^2)L [1/l_P+1/l_D]$. This resistance, if one
omits the drag contribution, is expressed in terms of the usual Drude
conductivity $\sigma= L/R=e^2 l_P/\pi \hbar= e^2 n \tau_P/m$. The
corresponding transresistance is

        \begin{equation}
        R_{TR}=-{\pi \hbar\over e^2}(L/l_D)\left[1 - (L/
        L_0)^2\right]$,~~ $L_0=\left(6
        l_P^2 l_T/l_D\right)^{1/2}.
        %31
        \end{equation}
Expressing $l_D$ and $l_T$ through the drag transresistivity
$\pi \hbar/e^2 l_D$ and the tunneling conductance $G_T=e^2
\rho_{1D}/\tau_T = 2 e^2/\pi \hbar l_T$, where $\rho_{1D}$ is
the 1D density of states at the Fermi level, one can see that Eq. (31)
formally coinsides with that obtained in Ref. 27, where a competition
of drag and tunneling effects was investigated for double quantum-well
systems. For $\lambda L \sim 1$, the transresistance is large and
comparable
to the direct resistance, because a considerable fraction of the current
penetrates the $l$ layer due to tunneling. This regime for double quantum
wells has been investigated both experimentally$^{28}$ and
theoretically$^{29}$. If one neglects the drag and assumes the diffusive
regime ($L \gg l_P$) with weak tunneling ($l_P \ll l_T$),
Eqs. (29) and (30) describe the results obtained in Ref. 29.
For $\lambda L \gg 1$  we have $R_{TR}=R=(\pi \hbar/e^2 l_P)(L/2)=
L/2 \sigma$. This is the case when the current, though injected only
in one layer, is equally distributed among the layers due to tunneling.

Figure 3 shows the length dependence of the transresistance calculated
for different relative contributions of the Coulomb drag and tunneling.
The transresistance is  negative for small $L$ but  always changes
its sign and becomes positive as $L$ increases and the backscattering
occurs often [see also Eq. (31)]. This behavior can be explained
with the help of the balance equation (28), which shows that the tunneling
tends to decrease the difference between $\mu_l^{\pm}$ and $\mu_r^{\pm}$
while the backscattering tends to decrease the difference between
$\mu_{l,r}^+$ and $\mu_{l,r}^-$. Thus, for
$\mu_l^+(0)=\mu_l^-(0)$ and $\mu_l^+(L)=\mu_l^-(L)$
the change of $\mu_l^+$ ($\mu_l^-$), with $x$, is opposite to that of
$\mu_r^+$ ($\mu_r^-$) at small $L$ and becomes the same as that of
$\mu_r^+$ ($\mu_r^-$) as $L$ increases, leading to the change of $R_{TR}=
[\mu_l^{\pm}(0)-\mu_l^{\pm}(L)] /eJ_r$ from negative to positive.
This transition occurs at smaller $L/l_P$ if the tunneling is stronger
(larger $l_P/l_T$) and the drag   weaker (smaller $l_P/l_D$).

Although $l_T$ is normally longer than $l_P$, the opposite
condition  can also be realized. A particularly interesting transport
regime, corresponding to long quantum wires without backscattering,
occurs in tunnel-coupled magnetic edge states$^{30,31}$, since an
edge state represents a 1D system where the electrons can move
only in one direction. Assuming $1/l_P=1/l_D=0$ in Eqs. (29) and
(30), we obtain the result of Ref. 31 in the form

        \begin{equation}
        R = \frac{\pi \hbar}{e^2}
        \left[ 1 + (1/2)\tanh ( L/l_T) \right],~~
        R_{TR} = -\frac{\pi \hbar}{2e^2} \tanh (L/l_T).
        %32
        \end{equation}

Consider now the behavior of the tunneling resistances. With Eqs. (26) and
(27) we obtain

        \begin{equation}
        R_{Ts} = {\pi \hbar\over 2e^2}\left[ 1+ \left(1+
        l_T/l_1 \right)^{1/2}
        \coth (\lambda L/2) \right]
        %33
        \end{equation}
and
        \begin{equation}
        R_{Tn} = {\pi \hbar\over 2e^2}\left[ 2 + L/ l_P + \left( 1+
        l_T /l_1 \right)^{1/2}
        \coth (\lambda L/2) \right],
        %34
        \end{equation}
for symmetric and non-symmetric contacts, respectively. For %c In
conditions $\lambda L/2 \ll 1$ we have $R_{Ts} \simeq R_{Tn} \simeq
(\pi \hbar/2 e^2) (l_T /L)$, i.e., the tunneling resistances
depend only on the ratio of the tunneling length to the wire length.
This is because the regime of $\lambda L/2 \ll 1$ corresponds to
weak tunneling and the chemical potentials $\mu_l^{\pm}(x)$ and
$\mu_r^{\pm}(x)$ are close to $\mu_{1l}$ and $\mu_{2r}$, respectively.
With the use of the tunneling conductance $G_T$ (see above) one can
rewrite the expression for the tunneling resistances in a more %c the most
transparent way: $R_{Ts} \simeq R_{Tn} \simeq (G_T L)^{-1}$.
For $\lambda L/2 \sim 1$, when the coordinate dependence of the chemical
potentials in the layers is essential, $R_{Tn}$ is different from
$R_{Ts}$ and both of them depend on the scattering length $l_P$. The
drag effect is not so important as for the transresistance: the
tunneling resistances depend on $l_D$ only if $l_D$ is comparable
to or smaller than both $l_P$ and $l_T$.

Figure 4 shows the length dependence of the tunneling resistances
$R_{Ts}$ and $R_{Tn}$, as given by Eqs. (33) and (34), for several
different values of the ratio $l_P/l_T$ describing the strength of the
tunneling with respect to the backscattering. The drag effect   %c s are
is neglected, $1/l_D =0$. As the wire length
becomes larger than the backscattering
length, the $1/L$ decrease of the tunneling resistance  changes to either
a $L$-independent behavior (for $R_{Ts}$) or to a linear increase %growth
(for $R_{Tn}$). In the first case the dependence on $L$ disappears
because all  tunneling occurs near the ends. In contrast, for
non-symmetric contacts the resistance $R_{Tn}$ is determined by
the Ohmic resistance of the wires instead of the tunneling effects, and
increases linearly with $L$. A similar effect, in applications to coupled
quantum wells, is discussed in Ref. 29.

\subsection{Short wires, $L \ll l_P, l_D$}

This length range  comprises the region from the purely ballistic ($l \ll
l_C$) to the "pseudo-ballistic" ($l_C \ll L \ll l_P, l_D$) regimes. Since
the electrons pass along the wires almost without backscattering, $R$ is
close
to $\pi \hbar/e^2$, and $R_{TR}$ is small. However, for $L \sim l_C$ an
electron tunneling between the layers does not lose its phase memory
completely and tunnel coherence effects can manifest themselves on
such short lengths giving additional contributions to the
transresistance $R_{TR}$ and the tunneling resistances $R_{Ts}$ and
$R_{Tn}$; accordingly the expressions for these quantities obtained in the
previous subsection for $L \ll l_P, l_D$ should be modified.

A convenient analytical approach to the problem in this regime is
to solve Eqs. (14)-(17) by iterations taking
$\mu_{l,r}^+(x)=\mu_{1 l,r}$, $\mu_{l,r}^-(x)=\mu_{2 l,r}$ and
$\mu_{u,v}^{\pm}(x)=0$ as an initial approximation.
Another way is to use Eqs. (19)-(22) directly. We obtain

        \begin{eqnarray}
        2 \mu_{l,r}^+(x) = \left(\mu_{1l}+\mu_{1r}\right)
        \left(1 - x/l_P \right) +
        \left(\mu_{2l}+\mu_{2r}\right)
        x/l_P \nonumber \\
        \pm \left(\mu_{1l}-\mu_{1r}\right)\left[ 1 - x/l_1
        - 2 x/l_T +2 (l_C/l_T) \Phi(x)  \right]
        \pm \left(\mu_{2l}-\mu_{2r}\right)
        x/l_1  \\
        \nonumber \\
        2 \mu_{l,r}^-(L-x) = \left(\mu_{2l}+\mu_{2r}\right)
        \left(1 -x/l_P \right) +
        \left(\mu_{1l}+\mu_{1r}\right)
        x/l_P \nonumber \\
        \pm \left(\mu_{2l}-\mu_{2r}\right) 
        \left[ 1 - x/l_1
        - 2 x/l_T +2 (l_C/l_T) \Phi(x) \right]
        \pm \left(\mu_{1l}-\mu_{1r}\right)
        x/l_1 %\nonumber \\
        %35,36
        \end{eqnarray}
where
        \begin{equation}
        \Phi(x)=[1/(l_C^{-2}+\delta_F^2)] \left[ 2 (\delta_F/l_C)
        e^{-x/l_C} \sin (\delta_F x)  %\nonumber  \right. \\
        + (l_C^{-2}-\delta_F^2)
        \left(1-e^{-x/l_C} \cos (\delta_F x) \right) \right],
        %37
        \end{equation}
%A
 is an oscillating function of the coordinate $x$ and
$\delta_F =\Delta /\hbar v_F$.
%A 
Now $R$ and  $R_{TR}$ are given, respectively, by

        \begin{equation}
        R = \frac{\hbar\pi}{e^2} \left[ 1 +L/l_P  + L/l_D +
          L/2l_T - (l_C/2 l_T) \Phi(L) \right],
        \end{equation}
and
        \begin{equation}
        R_{TR} = \frac{\hbar\pi}{e^2} \left[ -L/l_D -
          L/2l_T + (l_C/2 l_T) \Phi(L) \right],
        %38,39
        \end{equation}
The contribution to $R_{TR}$ coming from the term proportional to
$\Phi(L)$ is not small for $L \sim l_C$. It describes oscillations
damped due to the factor $\exp(-L/l_C)$.
The periodic behavior can be described as a result of the interference
of electron waves of the left and right layers along the %c coupling
length $L$: due to a finite level splitting $\Delta$ these waves have
different phase velocities.

Similar interference effects occur in the tunneling resistances
        \begin{equation}
        R_{Ts} \simeq R_{Ta} \simeq \frac{\hbar\pi}{2 e^2} (l_T/L)
        \left[ 1 - (l_C/L) \Phi(L) \right]^{-1}.
        %40
        \end{equation}
%A
%%Figure 4 shows the dependence of the factor $\Phi(L)$ entering Eqs.
%%(38)-(40) as a function of $\delta_F L=\Delta L/\hbar v_F$. In the
%%calculations we used the result (A14) for $\tau_C$ and $T \ll |\Delta|$,
%%so that $l_C^{-1}= \kappa |\delta_F|$, where $\kappa$ is a dimensionless
%%parameter.
From Eq. (37) for $\Phi(L)$ we see that both the transresistance and 
the tunneling resistances, being functions of $\Delta L/\hbar v_F$, oscillate with level 
separation $\Delta$. The 
oscillations are damped when the wire length $L$ exceeds $l_C$ 
so that the tunnel coherence over the wire length is suppressed.
%A

Changing $\Delta$ by applying a voltage across the wires would
lead to oscillations of $R$, $R_{TR}$, $R_{Ts}$, and $R_{Tn}$.
Another way to change $\Delta$ is to apply a magnetic field $B$
perpendicular to the plane of the wires$^{15,16}$. For sufficiently
weak $B$ the results presented so far still hold with the phase
$\delta_F L$ having an additional contribution  $2 \pi \phi/\phi_0$,
where $\phi_0=h/e$ is the magnetic flux quantum and $\phi = B w L$ the
flux enclosed by the area between the wires. Though the double-wire
system does not form a closed current loop, this should lead to
Aharonov-Bohm-type oscillations in the resistances defined by
Eqs. (38)-(40).

In very short wires, with $L \ll l_C, \delta_F^{-1}$, Eqs. (39) and (40)
become

\begin{equation}
R_{TR} = -{\pi \hbar\over e^2}\left(L  /l_D + L^2 t_F^2/
2\right),~~~
R_{Ts}=R_{Tn}= {\pi \hbar\over 2 e^2} L^{-2} t_F^{-2}.
%41
\end{equation}
In this regime only a small fraction of the electronic wave packet
is coherently transmitted from one wire to another. The tunneling
contribution to $R_{TR}$ follows a $L^2$ dependence, instead of the linear
dependence  occuring  for $l_C \ll L \ll \l_P,l_D$, when the tunneling
is non-coherent. The length dependence of the tunneling resistance
follows a $L^{-2}$ law.

In the investigation of the purely ballistic regime ($L \ll l_C$)
we can neglect the collision integral in Eq. (10) and
need not make the assumption about the smallness of the tunneling matrix
element which was essential for evaluation of the scattering-induced
contributions in Eqs. (14)-(17). The electron transport in coupled quantum
wires in this regime is pertinent to the problem of electron-wave
directional couplers. Theoretical studies of this problem$^{4-14}$,
although rather extensive, included only a
quantum-mechanical calculation of the electronic transmission.
Below we show how the essential results of these  studies can
be obtained in a simple way from the quantum-kinetic analysis.
Integrating Eq. (10), with $\delta \hat{I}_{\pm}(\varepsilon,x)=0$,
over the energy and taking Eq. (13) into account, we
find that the distribution of the chemical potentials
is again described by Eqs. (14)-(17) without the terms containing
the scattering lengths $l_C$, $l_P$, and $l_D$. Since there is no
backscattering, the solutions for $\mu_{l,r}^+$ and $\mu_{l,r}^-$
are decoupled

        \begin{eqnarray}
        \mu_{l,r}^+(x) = \mu_{1l,r} \mp (\mu_{1l} - \mu_{1r})
        r \sin^2(\Delta_T x/2 \hbar v_F), \\
        \mu_{l,r}^-(x) = \mu_{2l,r} \mp (\mu_{2l} - \mu_{2r})
        r \sin^2(\Delta_T (L-x)/2 \hbar v_F);
        %42,43
        \end{eqnarray}
here $\Delta_T=(\Delta^2 + 4T^2)^{1/2}$ and $r=4T^2/\Delta_T^2$.
Equations (42) and (43) describe oscillations of the electronic wave packets
between the layers due to coherent tunneling. A complete transfer of the
wave packet can be achieved for $\Delta=0$. One can calculate
the resistance and transresistance as

        \begin{eqnarray} 
        R = {\pi \hbar\over  e^2} [ 1  - (r/2) \sin^2 \psi]/
        [ 1 - r \sin^2 \psi ], \\
        R_{TR} = -{\pi \hbar\over 2 e^2} r \sin^2 \psi/
        [ 1 - r \sin^2 \psi ]  %^{-1},
        %44,45
        \end{eqnarray}
and the tunneling resistances as

        \begin{eqnarray}
        R_{Ts} = R_{Tn} - {\pi \hbar\over 2 e^2}=
        {\pi \hbar\over 2 e^2} r^{-1} \sin^{-2} \psi,~~
        %46
        \end{eqnarray}
Here $\psi=\Delta_T L/2 \hbar v_F$.
The oscillations of these quantities occur in a
way similar to the one described by Eqs. (38)-(40): $\delta_F L$
coinsides with $2 \psi$ if
one replaces $\Delta$ by $\Delta_T$. However, since the tunnel coupling
is strong, the oscillations described by Eqs. (44)-(46) have large
amplitudes. In particular, when $\Delta$ is small 
($r \simeq 1$), all the quantities
given by Eqs. (44)-(46) show giant oscillations with amplitude
large in comparison to $G_0^{-1}$.

\section{Conclusions}

In this paper we carried out a theoretical study of electron
transport in parallel 1D layers coupled by tunneling and Coulomb
interaction and contacted, at their ends, to quasi-equilibrium
reservoirs. A linear-response, steady-state regime has been
investigated, and the wires were assumed to be ideal,
i.e., without defects and, therefore, the elastic scattering of
electrons by them was neglected. As the most important result
of our study, we found that a full quantum-kinetic description
of the problem is reduced, with physically reasonable
assumptions, to a set of linear, first-order differential equations
describing the distribution of local chemical potentials
for forward- and backward-moving electrons. The boundary conditions
for the chemical potentials are determined by the potentials of the
reservoirs controlled by applied voltages. The solution of this set
was obtained analytically and allowed us to describe the local
currents flowing in each layer from the pure ballistic regime,
when the electrons do not suffer any scattering along the wires,
to the diffusive regime, when the electrons
experience many backscattering events during the transport.

In particular, we applied our approach to the description of the
resistance $R$, transresistance $R_{TR}$, and tunneling resistances
$R_{Ts}$ and $R_{Tn}$ of double quantum wires. The most
important result is that $R_{TR}$, which is caused by both
tunneling and Coulomb drag effects, depends on the wire length $L$
non-monotonically and always changes its sign
as  $L$ increases,  because in shorter wires,
when backscattering is rare, the tunneling, as well as the drag,
leads to a {\em negative} $R_{TR}$, while in longer wires,
when the transport becomes diffusive, the tunneling
leads to a {\em positive} $R_{TR}$, in a way similar to that for coupled
two-dimensional (2D) systems$^{28,29}$, and overcomes the drag as
the length $L$ increases. This sign inversion  is qualitatively
understood from an analysis of the balance equation (28) and
is mathematically described by Eq. (30). In the diffusive limit
and for $l_T \gg l_P$, Eqs.  (29) and (30) formally coinside with those
obtained previously$^{27,29}$ for coupled 2D
systems. Besides, some recent studies of transport phenomena in
coupled 1D layers, namely transport without backscattering in
tunnel-coupled edge states$^{30,31}$ and Coulomb drag between
quantum wires in the ballistic regime$^{21}$, constitute limiting
cases of the more general  results given by Eqs. (29) and (30).

One should stress the importance of the phase-breaking processes
that suppress the tunnel coherence. In our model, i.e., without
elastic scattering, these processes are proved to be much more frequent
than the backscattering processes. This
allowed us to distinguish two transport regimes: the pure ballistic
regime, without any scattering, and the pseudo-ballistic one, without
backscattering but with essential
forward-scattering due to electron-electron interaction, and
with suppressed coherence. If one considers just a single wire, there
is no difference between these regimes as concerns the electrical
properties: the resistance is equal to $G_0^{-1}$ in both cases.
However, the electrical properties of a tunnel-coupled double-wire
system behave differently as one passes from one regime to another,
because the contribution of the tunneling to the electrical properties
becomes different. In the ballistic regime, as well as in the transition
region between the two regimes ($L \sim l_C$), all calculated resistances
oscillate with $L$ and with the level splitting energy $\Delta$ due
to interference of the electron waves. One can vary the level
splitting energy by applying either a transverse voltage across the
wires or a magnetic field perpendicular to the wire plane. In the
latter case the oscillations show a Aharonov-Bohm periodicity associated
with the magnetic flux penetrating through the area $L w$ between the
wires. The oscillations become exponentially damped as the ratio $L/l_C$
increases. If the tunnel coupling is strong, the oscillations
have large amplitudes, which, from a theoretical point of view,
can be much larger than the resistance quantum $G_0^{-1}$.
So far the experimentally observed$^{9,12}$ resistance oscillations in
tunnel-coupled ballistic quantum wires were of small ($\sim$ 0.5
K$\Omega$) amplitude. This is not surprising because there are
many factors which compete against the tunnel coherence. Apart
from inelastic scattering considered in this paper, there are
elastic scattering and long-scale inhomogeneities of the wires
which would lead to a coordinate dependence of the level splitting
$\Delta$. If these variations of $\Delta$ are larger than the
tunneling matrix element, the coherence would be considerably suppressed.

We now discuss the approximations made in this paper. The main
approximation is the neglect of elastic scattering. Since this
scattering tends to be dominant at low temperatures, the presence
of impurities in the 1D channels will considerably modify the
transport. The elastic scattering will lead to an increase of
backscattering and interference between the forward- and backward-moving
electron waves. As a consequence, $R$, $R_{TR}$, $R_{Ts}$ and $R_{Tn}$
will depend on the spatial positions of the impurities in the channels
and the regular expressions obtained in this paper will not be valid.
A further development of the transport theory for tunnel-coupled
wires in the presence of elastic scattering is therefore desirable.
On the other hand, advances in the technology of nanostructures, in
particular selective doping, can make it possible to achieve
structures where the elastic scattering in 1D channels is minimized
for wire lengths smaller than a few microns which is the current
standard of the impurity mean free path at low temperatures.

Another approximation concerns the transition from the quantum kinetic
equation (1) to the semi-classical description given by Eq. (7). It is
valid when the spatial scale of the electronic distribution is large
in comparison to the Fermi wavelength $\pi \hbar/p_F$. We have seen
that this scale is determined either by $\lambda$, given by Eq. (24)
for long wires, or by $\lambda_{\pm}$ for short wires.
In the case of strong tunnel coupling the characteristic scale is given
by $\hbar v_F/\Delta_T$. Therefore, the necessary requirement is
fulfilled if the tunneling matrix element $T$, level splitting $\Delta$,
and the energy $\hbar/\tau_C$, associated with the smallest scattering
time $\tau_C$, are small in comparison to the Fermi energy. These conditions
have been assumed througout the paper. This also allowed us to neglect
the difference between the electron densities in the layers and
characterize the electrons in different layers by the same Fermi
velocity $|v_p| \simeq v_F$.

The assumption about the adiabatic connection of the wires to the leads,
which allowed us to neglect elastic scattering of electrons near the
ends of the wires, implies that the Fermi wavelength $\pi \hbar/p_F$
must be small in comparison to the contact lengths, i.e., to the lengths
of transition from the leads to the wires. On the other hand, the
oscillations associated with the tunnel coherence,  cf.
Sec. III B, can be seen if the contact lengths are smaller than
both $l_C$ and $\hbar v_F/\Delta_T$. In principle, both
requirements can be fulfilled.

The next approximation, which allowed us to solve the kinetic equation
analytically in the whole range of regimes from ballistic to diffusive,
is equivalent to the following statement. In each layer the
forward- and backward-moving electrons can be described as
weakly coupled sub-systems characterized by their own local chemical
potentials. This statement is obvious for the case of pure
ballistic or pseudo-ballistic transport, when these potentials are
merely dictated by the reservoirs (leads) and do not change with
coordinate $x$. When the backscattering becomes essential, this
statement is still true if we assume that the forward-scattering
events are much more frequent than the backscattering and tunneling
events. For example, it is always true for magnetic edge
states, where one can completely neglect backscattering, and
the introduction of local chemical potentials (see Ref. 31) is
well-justified.  In our case, a consideration of the
electron-electron collision integral allowed us to estimate the
characteristic time of the Coulomb-assisted forward scattering,
and we find that it is of the order of $\tau_C$, which is small in
comparison with both backscattering times $\tau_P$ and $\tau_D$.
Thus, the electron-electron interaction provides an effective
mechanism for forward scattering and can maintain quasi-equilibrium
Fermi distribution functions for  forward- and backward-moving
electron sub-systems. However, these conditions may be violated
when the conducting channels contain impurities with short-range
potentials and the elastic backscattering becomes important.

%A
Our evaluation of the characteristic scattering times from the
collision integral has employed only the lowest-order essential
contributions of the electron-phonon and electron-electron 
interactions, given by the diagrams of 
Fig. 2 and leading to collision integrals with scattering 
amplitudes in the Born approximation. While it is normally$^{32}$ 
good for electron-phonon
interaction due to the weakness of the coupling constant, a rigorous
evaluation of the electron-electron part requires also a consideration
of higher-order contributions, given by more complex diagrams, because
the ratio of the Bohr energy to Fermi energy $\varepsilon_F$, which is
the parameter of the perturbation expansion for the Coulomb interaction,
is not small. 
%A
Nevertheless,  using the Born approximation 
in the evaluation of the drag time is still reasonable if the momentum $2p_F$
transferred in backscattering is large and the electron-electron 
backscattering probability is small. As concerns $\tau_C$, it is 
determined by forward-scattering processes with small momentum 
transfer and the Born approximation is not well justified$^{33}$. On the 
other hand, our theory leads to a non-divergent expression (A14) for 
$\tau_C$ and gives, for typical parameters of the electron system, 
physically reasonable values. We remind that in our theory both 
$k_B T_e$ and $\Delta$ are much smaller than $\varepsilon_F$. 
Therefore, one may expect that Eq. (A14) provides a correct 
order-of-magnitude estimate of the phase-breaking time 
caused by electron-electron interaction.

%%On the other hand, for $\varepsilon_F \sim 5$ meV, this
%%parameter is  not big and we may claim that the calculation of
%%$\tau_C$ done in this paper gives us a correct order-of-magnitude
%%estimate of the phase-breaking time caused by electron-electron interaction.

Finally, we stress that the results obtained in this paper hold for 
a normal Fermi-liquid state of the electron system. If the electrons in 
the wires are in the Luttinger-liquid state$^{34}$, these results have to 
be reconsidered.  
%A

\acknowledgements

This work was supported by the Canadian NSERC Grant No. OGP0121756.

{{\center \bf APPENDIX\\}}

Below we give a microscopic calculation of the characteristic
times $\tau_P$, $\tau_D$, and $\tau_C$. The coordinate index $x$ in
the Green's functions and self-energies is omitted
and $\hbar$ is set equal to 1. 
%A
The normalization lengths 
are also set equal to 1. 
%A
The electron-phonon self-energies
given by the diagrams shown in Fig. 2 (a) are explicitly expressed as

        $$
        \Sigma^{\alpha \beta}_{jj', {\textstyle \varepsilon}}(p)=
        i \sum_{{\bf Q}} \int \frac{d \omega}{2 \pi}
        G^{\alpha \beta}_{jj', {\textstyle \varepsilon-\omega}}(p-q)
        D^{\alpha \beta}_{jj'}(\omega, Q) M^{e-ph}_{jj}({\bf Q})
        M^{e-ph}_{j'j'}(-{\bf Q}),
        \eqno(A1)
        $$
where the unperturbed Green's functions of phonons $D^{-+}$ and $D^{+-}$
(we do not need $D^{--}$ and $D^{++}$ in the following) are given as

        $$
        i D^{\mp \pm}_{jj'}(\omega, Q) =
        2 \pi \left[ N_Q \delta(\omega \mp sQ) +
        (1+N_Q) \delta(\omega \pm sQ) \right]
        \eqno(A2)
        $$
and the matrix elements of the electron-phonon interaction are

        $$
        M^{e-ph}_{jj}({\bf Q})= i \sqrt{ E_1^2 Q/2 \rho s }
        J^{e-ph}_j(q_y,q_z),
        $$
        $$
        J^{e-ph}_j(q_y,q_z)=
        \int \int dy dz F_j^2(y,z) e^{i q_y y + iq_z z}.
        \eqno(A3)
        $$
We use the expression $sQ$ for the phonon energy, where
$Q=|{\bf Q}|$, ${\bf Q}=(q,q_y,q_z)$ is the phonon wave vector and $s$
 the velocity of sound. Further, $N_Q=1/[\exp(sQ/k_{B}T_e)-1]$ is the
Planck distribution function, $\rho$ is the material density, and
$E_1$ is the deformation potential constant.

The electron-electron self-energies given by the diagrams shown in
Fig. 2 (b) are expressed as

        $$
        \Sigma^{\alpha \beta}_{jj' {\textstyle \varepsilon}}(p) = 2 (-1)^l
        \sum_{j_1 j'_1} \sum_{p',q} M^{e-e}_{jj_1}(q) M^{e-e}_{j'j'_1}(-q)
        $$
        $$
        \times \int \int \frac{ d \omega d \varepsilon'}{(2 \pi)^2}
        G^{\alpha \beta}_{jj', {\textstyle \varepsilon-\omega}}(p-q)
        G^{\beta \alpha}_{j'_1 j_1, {\textstyle \varepsilon'}}(p')
        G^{\alpha \beta}_{j_1 j'_1, {\textstyle \varepsilon'+
        \omega}}(p'+q),
        \eqno (A4)
        $$
where $l=0$ for $\alpha=\beta$ and $l=1$ for $\alpha \neq \beta$, the
factor of 2 comes from the spin summation in the "loop". Here

        $$
        M^{e-e}_{jj'}(q)= (2e^2/\epsilon) \int \int  \int \int d y d y'
        dz dz' K_0(|q||{\bf r}- {\bf r}'| ) F_j^2(y,z) F_{j'}^2(y',z')
        \eqno (A5)
        $$
are the matrix elements for electron-electron interaction, $\epsilon$
is the dielectric constant, $K_0$ is the modified Bessel function, and
$|{\bf r}- {\bf r}'|= [(y-y')^2+(z-z')^2]^{1/2}$.

For the evaluation of the collision integral we use Eqs. (3)-(6)
and express the non-equilibrium part of the matrix Green's functions
according to [see also Eq. (9)]

$$
\delta \hat{G}^{\alpha \beta}_{{\textstyle \varepsilon}}(\pm|p|)=
\hat{g}^{\pm}_{\textstyle \varepsilon}
\hat{G}^A_{{\textstyle \varepsilon}}(p) -
\hat{G}^R_{{\textstyle \varepsilon}}(p)
\hat{g}^{\pm}_{\textstyle \varepsilon}.
\eqno (A6)
$$

The collision integrals are evaluated below assuming %c in approximation
of weak  tunneling, when the nondiagonal contributions of
$\hat{G}^{R,A}_{{\textstyle
\varepsilon}}(p)$ are neglected. This approximation is valid when the
tunneling matrix element $T$ is small in comparison to the imaginary
part of the self-energies and when the level splitting $|\Delta|$ is
small in comparison to the Fermi energy. Both
requirements are assumed fulfilled. Then the components
$[\hat{g}^{\pm}_{\textstyle
\varepsilon}]_{jj'}$ enter only in the corresponding parts
$\delta [\hat{I}(\varepsilon)]_{jj'}$ of the collision integral.

Now we calculate the {\it diagonal} parts for the electron-phonon
scattering contribution to the collision integral.
Taking the self-energy given
by Eqs. (A1)-(A3) we find $\int d \varepsilon \delta
[\hat{I}^{e-ph}_{+}(\varepsilon)]_{jj} = -(\mu_j^+ -\mu_j^-)/2\tau_{Pj}$
where $j=l,r$.  The phonon-assisted transport time  is given by

        $$
        \tau_{Pj}^{-1}= \frac{E_1^2}{\rho s T_e} \sum_{q_y,q_z}
        |J^{e-ph}_j(q_y,q_z)|^2
        \sum_{p, q (p>0, p-q <0)} v_p Q \int d \varepsilon
        \frac{[f(\varepsilon)-f(\varepsilon-sQ)]}{4 \sinh^2(sQ/2T_e)}
        $$
        $$
        \times\left[G^c_{jj, {\textstyle \varepsilon-sQ}}(p)
        G^c_{jj, {\textstyle \varepsilon}}(p-q) +
        G^c_{jj, {\textstyle \varepsilon}}(p)
        G^c_{jj, {\textstyle \varepsilon-sQ}}(p-q) \right],
        \eqno(A7)
        $$
where we defined $G^c=G^A-G^R$. For further evaluation of $\tau_{Pj}$
we use the free-particle (unperturbed) Green's functions, i.e.,
$G^{R,A}_{ll, {\textstyle \varepsilon}}(p) = [\varepsilon - \Delta/2 -
p^2/2m \pm i0]^{-1}$ and $G^{R,A}_{rr, {\textstyle \varepsilon}}(p)
= [\varepsilon + \Delta/2 - p^2/2m \pm i0]^{-1}$, and obtain

        $$
        \tau_{Pj}^{-1} = \frac{E_1^2}{\rho s k_{B}T_e} \sum_{q_y,q_z}
        |J^{e-ph}_j(q_y,q_z)|^2  \int_{-\infty}^{\infty} d q_1
        \frac{Q[f( \mu + v_F q_1/2-sQ/2)-f(\mu + v_F q_1/2+sQ/2)]}{
        4 \sinh^2(sQ/2k_{B}T_e)},
        \eqno(A8)
        $$
where $q_1= q - 2 p_F$ is a small variable. In the limit $k_{B}T_e \gg m
s^2$ the integral over $q_1$ is easily calculated. In addition, if $k_{B}T_e
\gg s [(\pi/a)^2+ (2 p_F)^2]^{1/2}$, where $a$ is the wire width, the
scattering becomes quasi-elastic, and Eq. (A8) is reduced to a
known$^{26}$ result

        $$
        \tau_{Pj}^{-1}= \frac{2 E_1^2 k_{B}T_e}{\rho s^2 v_F}
        \int \int dy dz
        F_j^4(y,z).
        \eqno(A9)
        $$
Since we assume that the confining potentials for the layers are almost
identical, the difference between $\tau_{Pl}$ and $\tau_{Pr}$ is
neglected: $\tau_{Pl}=\tau_{Pl}=\tau_P$. A numerical estimate, using Eq.
(A9) and GaAs material parameters, gives
$\tau_P^{-1} \sim$ 10$^{-2}$ $k_{B}T_e$.

The electron-electron scattering contribution to the {\it diagonal}
parts of the collision integrals gives the Coulomb drag terms:
$\int d \varepsilon \delta [\hat{I}^{e-e}_{\pm}(\varepsilon)]_{jj} =
-(\mu_j^{\pm} - \mu_j^{\mp})/2\tau_D +(\mu_{j'}^{\pm} -\mu_{j'}^{\mp})/
2\tau_D$, where $j' \neq j$. The drag time is given by

        $$
        \tau_D^{-1}= \frac{4}{k_B T_e} \sum_{p,p',q} |M^{e-e}_{lr}(q)|^2 v_p
        \int \int
        \int \frac{ d \varepsilon d \varepsilon' d \omega}{(2 \pi)^2}
        G^c_{ll, {\textstyle \varepsilon}}(p)
        G^c_{ll, {\textstyle \varepsilon-\omega}}(p-q)
        $$
        $$
        \times G^c_{rr, {\textstyle \varepsilon'}}(p')
        G^c_{rr, {\textstyle \varepsilon'+\omega}}(p'+q)
        \frac{[f(\varepsilon-\omega)-f(\varepsilon)]
        [f(\varepsilon')-f(\varepsilon'+\omega)]}{4
        \sinh^2(\omega/2k_{B}T_e)}
        \eqno(A10)
        $$
The sum here must be evaluated for  $p>0$, $p-q<0$, $p'<0$, and $p'+q >0$.
The evaluation of   Eq. (A10)
using the free-particle Green's functions gives a simple result

        $$
        \tau_D^{-1}= \frac{k_{B}T_e}{\pi v_F^2} |M^{e-e}_{lr}(2 p_F)|^2
        \frac{(\Delta/2 k_{B}T_e)^2}{\sinh^2(\Delta/2k_{B}T_e)}
        \eqno(A11)
        $$
One can estimate $M^{e-e}_{lr}(2 p_F)$ as $(2 e^2 /\epsilon) K_0(2 p_F
w)$, where $w$ is the distance between the
centers of the wires. If $2 p_F w \ll 1$, which
is easily achieved for $w \sim 30$ nm, $K_0(2 p_F w)$ is exponentially
small.

Finally, we calculate the electron-electron part of the {\it nondiagonal}
components of the collision integral. Since the main contribution to it
comes from the forward-scattering processes ($|q| \ll p_F$), only
such processes  are considered below. The integral of
$[\hat{I}^{e-e}_{\pm} (\varepsilon)
]_{jj'}$ ($j \neq j'$) over the energy $\varepsilon$ can be reduced to a
sum of three terms characterized by three different statistical factors:

        $$
        \int d \varepsilon \delta [\hat{I}^{e-e}_{\pm}
        (\varepsilon)]_{jj'} =
        - 2 \int d \varepsilon [g^{\pm}_{\textstyle \varepsilon}]_{jj'}
        \sum_{p,p',q (p>0, p-q >0)}  v_p
        \int \int \frac{d \varepsilon' d \omega}{(2 \pi)^2}
        $$
        $$
        \times \left\{ \left(\Lambda^{ARAR}_{j'j} +
        \Lambda^{RARA}_{jj'} \right) \left[
        f(\varepsilon')[1-f(\varepsilon'+\omega)] +
        f(\varepsilon-\omega) [f(\varepsilon'+\omega)-f(\varepsilon')]
        \right] \right.
        $$
        $$
        + \left.  \left(\Lambda^{ARRA}_{j'j} + \Lambda^{RAAR}_{jj'}
        \right)
        f(\varepsilon'+\omega) [1-f(\varepsilon')] +
        \left(\Lambda^{ARRA}_{j'j'} + \Lambda^{RAAR}_{jj} \right)
        f(\varepsilon-\omega) [f(\varepsilon'+\omega)-f(\varepsilon')]
        \right\}.
        \eqno (A12)
        $$
In Eq. (A12) we used the shortcuts

        $$
        \Lambda^{\alpha \beta \gamma \delta}_{jj'}=
        G^{\alpha}_{jj,{\textstyle \varepsilon}}(p)
        G^{\beta}_{j'j',{\textstyle \varepsilon-\omega}}(p-q)
        \sum_i G^{\gamma}_{ii,{\textstyle \varepsilon'}}(p')
        G^{\delta}_{ii,{\textstyle \varepsilon'+\omega}}(p'+q)
        $$
        $$
        \times \left[ \left( M^{e-e}_{j'i}(q)\right)^2-M^{e-e}_{ji}(q)
        M^{e-e}_{j'i}(q) \right],
        \eqno (A13)
        $$
and neglected the terms with $\alpha=\beta$ and $\gamma=\delta$ because
they vanish after the summations over $p$ and $p'$, respectively.
Since $[g^{\pm}_{\textstyle \varepsilon}]_{lr} = [g^{\pm}_{\textstyle
\varepsilon}]_{rl}^* = g^{\pm}_{u, {\textstyle \varepsilon}} -
i g^{\pm}_{v, {\textstyle \varepsilon}}$, one can see that
$\int d \varepsilon \delta [\hat{I}^{e-e}_{\pm} (\varepsilon)]_{jj'} =
\int d \varepsilon \delta [\hat{I}^{e-e}_{\pm} (\varepsilon)]_{j'j}^*$.

Calculating the integrals in the expression (A12) within the approximation
of the free-particle Green's functions, we find that the third term
on the right-hand side of Eq. (A12) vanishes. The first term diverges
for $\Delta=0$ but it is close to zero for $\Delta \neq 0$ and can be
neglected in the following. In contrast, the second term gives a regular
contribution, which can be represented,
on account of Eq. (13), as
$\int d \varepsilon \delta [\hat{I}^{e-e}_{\pm} (\varepsilon)]_{jj'}
=-\mu^{\pm}_{jj'}/\tau_C$. The "nondiagonal" relaxation time $\tau_C$
(we take into account only its {\it real} part) %c of it)
is given by

        $$
        \tau_C^{-1}=\frac{e^4 S^2 \Delta}{2 \pi \epsilon^2 v_F^2}
        \coth \frac{\Delta}{4 k_{B}T_e},
        \eqno(A14)
        $$
where

        $$
        S = - \int \int \int \int dy dy' dz dz' \ln |{\bf r}- {\bf r}'|
        F_l^2(y,z) [F_l^2(y',z') - F_r^2(y',z') ].
        \eqno(A15)
        $$
In the calculation we took into account $M^{e-e}_{ll}(q)
\simeq M^{e-e}_{rr}(q)$ and $qa \ll 1$. The last property
allowed us to use the approximation
$K_0(x) \simeq -[C+\ln(x/2)]$, where $C$ is Euler's constant;
we found   $M^{e-e}_{ll}(q)-M^{e-e}_{lr}(q) \simeq
(2e^2/\epsilon) S$. The overlap integral $S$ can be approximated,
to a good accuracy, by $\ln (w/a)$.

If $\Delta \gg 4 k_{B}T_e$, the relaxation rate $\tau_C^{-1}$ given by Eq.
(A14) is temperature-independent and proportional to $|\Delta|$. For
$\Delta
\ll 4 k_{B}T_e$, $\tau_C^{-1}$ is proportional to $T_e$. A comparison of
Eq. (A11) and Eq. (A14) shows that $\tau_C$ is always much smaller than
$\tau_D$, since $\tau_C$ is controlled by  forward-scattering
processes and does not contain the smallness associated with the factor
$[K_0(2 p_F w)]^2$. A numerical estimate also shows that $\tau_C \ll
\tau_P$, because of the weakness of the electron coupling to acoustical
phonons. For this reason we neglected the contribution of electron-phonon
scattering to the nondiagonal part of the collision integral.

\clearpage

\begin{figure}

\caption{Schematic representation of two coupled quantum wires.}
\label{fig.1}
\ \\
\caption{Feynman diagrams describing the contributions of electron-phonon
(a) and electron-electron (b) interaction to the self-energies.}
\label{fig.2}
\ \\

\caption{Transresistance $R_{TR}$ as a function of the wire length $L$.
The solid curves correspond to $l_P/l_D=0.1$ (weaker drag) and the dashed
ones
to $l_P/l_D=1$ (stronger drag). Each curve is marked by the value of
$l_P/l_T$.}
\label{fig.3}
\ \\

\caption{Tunneling resistances for symmetric (solid) and non-symmetric
(dashed)  setups as a function of the wire length $L$, at $1/l_D=0$.
Each curve is marked by the value of $l_P/l_T$. The inset shows the
currents (arrows) injected in and coming out of the wires (broad lines)
for both  cases.}

\end{figure}

\end{document}